# Earthquakes, Hurricanes, and Mobile Communication Patterns in the New York Metro Area: Collective Behavior during Extreme Events

Christopher Small[1], Richard Becker[2], Ramón Cáceres[2,3], Simon Urbanek[2]

*Lamont-Doherty Earth Observatory[1]*  *AT&T Labs – Research[2]*
*Columbia University*  *Bedminster, NJ, and*
*Palisades, NY, USA*  *New York, NY, USA*

## Abstract

We use wireless voice-call and text-message volumes to quantify spatiotemporal communication patterns in the New York Metro area before, during, and after the Virginia earthquake and Hurricane Irene in 2011. The earthquake produces an instantaneous and pervasive increase in volume and a ~90-minute temporal disruption to both call and text volume patterns, but call volume anomalies are much larger. The magnitude of call volume anomaly diminishes with distance from earthquake epicenter, with multiple clusters of high response in Manhattan. The hurricane produces a two-day, spatially varying disruption to normal call and text volume patterns. In most coastal areas call volumes dropped anomalously in the afternoon before the hurricane's arrival, but text volumes showed a much less consistent pattern. These spatial patterns suggest partial, but not full, compliance with evacuation orders for low-lying areas.  By helping us understand how people behave in actual emergencies, wireless data patterns may assist network operators and emergency planners who want to provide the best possible services to the community.  We have been careful to preserve privacy throughout this work by using only anonymous and aggregate data.

## Introduction

Understanding dynamics of collective human behavior during extreme events has obvious relevance to both preparedness and response.  Most current knowledge of human behavior during extreme events comes from relatively small numbers of retrospective observations — often qualitative with unknown accuracy or degree of representation of the impacted population.  In contrast, mobile communication data can provide pervasive, quantitative observations of human communication patterns before, during and after extreme events.  With sufficient spatial and temporal detail, mobile communication data can even be used as proxies for other types of collective behavior (e.g. mobility (Becker et al. 2013; González et al. 2008)).  In this study we use a spatially and temporally extensive collection of voice call and SMS text message volumes to quantify spatiotemporal communication patterns in the New York Metro area before, during and after the magnitude 5.8 Virginia earthquake (2011-08-23) and the passage of Hurricane Irene (2011-08-28), both of which occurred in the same week of 2011. We compare and contrast spatial and temporal disruptions to normal patterns of voice and text communication in response to each of these extreme events in a diverse range of environments within the New York Metro area.

---

[3] Ramón Cáceres is now at Google, New York, NY, USA.

The focus of our analysis is on the spatiotemporal response of a large city and its surrounding suburban and rural communities to two distinct types of extreme event. The earthquake represents an unexpected abrupt event detected by almost everyone over a large area. The hurricane represents a forecast event with geographically varying impact over a large area. Differences in anticipation, impact and duration represent three important dimensions of extreme events and how they are both prepared for and responded to. In a comparison of mobile communication response to several different types of extreme event (blackout, bombing, earthquake, plane crash and storm) in a Western European country, (Bagrow et al. 2011) found that all these types of events cause communication spikes that are both spatially and temporally localized but that the spikes trigger communication avalanches that engage the social networks of eyewitnesses extending beyond the area impacted by the event. Our objective is to compare and contrast the spatial and temporal responses of both voice call and SMS text communications to two very different, and very unusual, cases of natural extreme event. Both hurricanes and earthquakes are extremely rare in the New York Metro area. Hurricane Irene was only the second hurricane (after an unnamed hurricane in 1821) to hit NYC directly in recorded history. The Virginia Earthquake (earthquake) of 2011 was the most widely felt (as reported) earthquake in US history and the largest to occur in the eastern US since a large (magnitude unknown) earthquake in 1897.

The mobile communication data used for this analysis have sufficient spatial and temporal resolution to characterize a wide range of responses as disruptions to normal communication patterns over a range of spatial and temporal scales. Numbers of voice calls (calls) and SMS text messages (texts) were measured at 1 minute intervals over the course of a year (2/1/2011 through 1/31/2012) at the spatial resolution of azimuthal sectors of almost 11,000 mobile network antennas within a 50 mile (80.5 km) radius of Times Square in NYC. We define a sector as the coverage area of the set of cellular antennas mounted in the same location (e.g., tower) and pointed in the same direction (e.g., north). We aggregate these call and text volumes at different spatial and temporal resolutions and use spatial correlation matrices to quantify normal spatial and temporal patterns and their disruption before, during and after both events. We have been careful to preserve privacy throughout this work. In particular, this study uses only the anonymous and aggregate data just described.

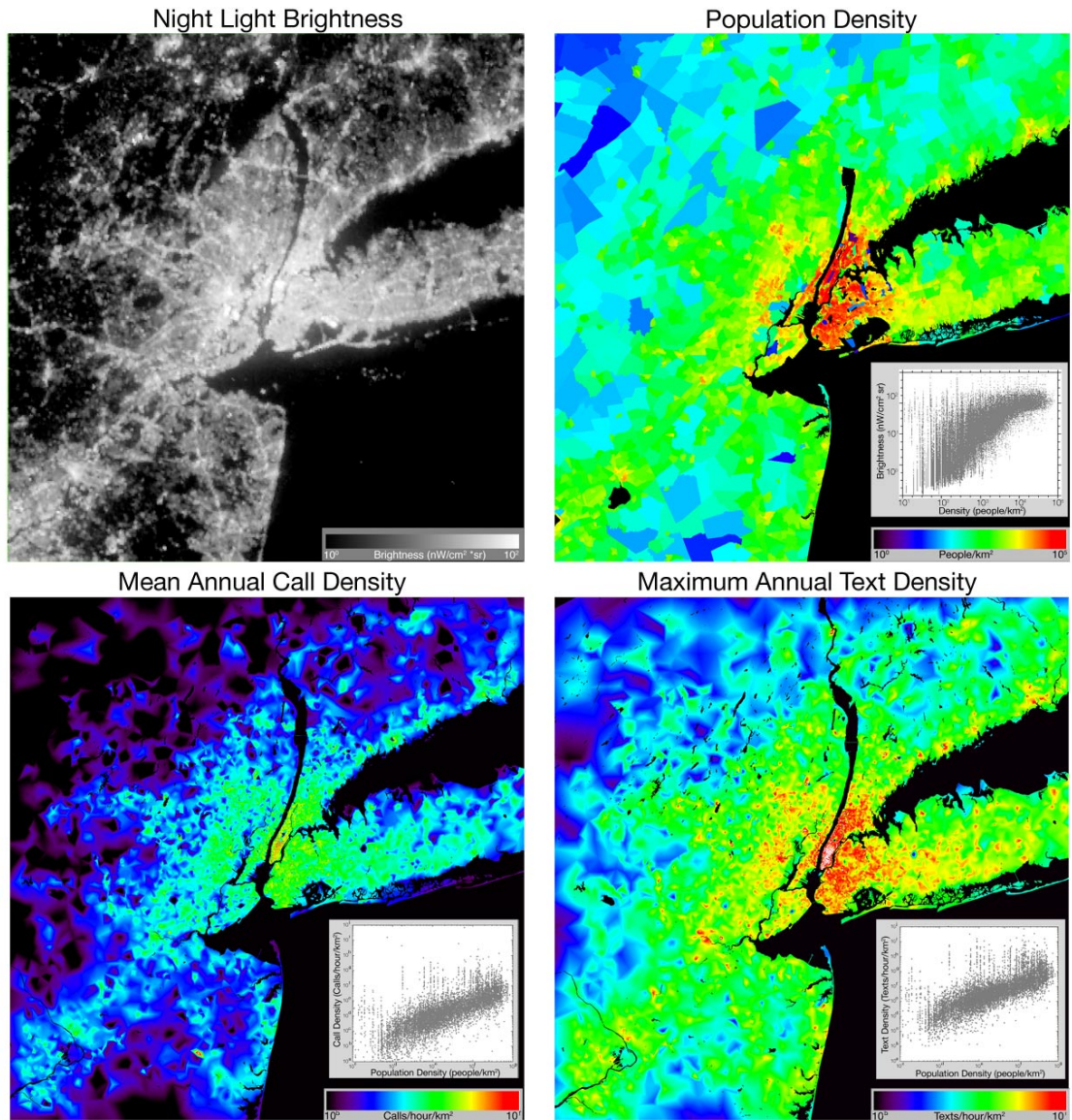

Figure 1 Spatial distribution of night light, population and mobile communication for the New York metro area. Night light brightness shows outdoor lighting - typically corresponding to intensity of development. Population density shows where people sleep. At the scale of the study area, mean annual call density and maximum annual text density have very similar spatial distributions, differing primarily in magnitude. Both call and text densities scale quasi-linearly with population density over 4 orders of magnitude. All quantities are displayed on logarithmic scales, showing the strong spatial gradients in each proxy for different aspects of human activity.

Some of the questions we seek to answer are:

1) What are the "normal" spatial and temporal patterns of communication in the NY Metro area?
2) What are the spatial and temporal scales of disruption to these patterns coinciding with the earthquake and hurricane?
3) What similarities and differences are observed for calls and texts under normal circumstances and during each event?

4) What inferences can be drawn from mobile communication data about collective behavior during extreme events?

The results of our analysis show both similarities and differences in call and text responses to both the earthquake and the hurricane. The earthquake produces a temporal disruption to both call and text volume patterns while the hurricane produces a spatial disruption to call and text volume patterns. We infer that the difference in the responses is a consequence of the fact that the hurricane was forecast days before its arrival allowing people in coastal areas time to relocate while the earthquake was an unexpected shock that prompted widespread instantaneous response throughout the study area. Comparison of call and text response to these events yields less intuitive results. While both call and text volumes increase abruptly following the earthquake, call volume anomalies are much larger than text volume anomalies but have a simpler impulse response and decay. On the day preceding the arrival of the hurricane, coastal evacuation zones show varying response both in location and in call versus text volume. In several evacuation zones call volumes decrease and text volumes increase throughout the day before the arrival of the hurricane.

## Results

### *Spatiotemporal Dynamics of Mobile Communication*

We begin by quantifying the normal spatial and temporal mobile communication patterns in the New York Metro area, then quantify the spatial and temporal disruptions coinciding with the earthquake and hurricane.

At the scale of the study area the spatial distribution of both calls and texts bears strong resemblance to both population density and intensity of development — but significant differences are observed at finer scales. Figure 1 shows annual mean call volumes and annual maximum text volumes compared to two complementary metrics for population and development. Gridded population density from the 2010 census shows the spatial distribution of where people reside. The census data represent the estimated number of residents within administrative units of census tracts. Census tracts are designed to scale approximately inversely with population density to provide greater spatial detail in areas of higher density. Night light brightness shows the spatial distribution of outdoor-lighted infrastructure as imaged by the day-night band of the Visible Infrared Radiometer Suite (VIIRS) on board the NASA/NOAA Suomi satellite. The sensor has an Instantaneous Field Of View (pixel spatial resolution) of ~500 m on the ground. The image in Fig. 1 represents a multitemporal average of emitted visible and Near Infrared radiance collected around 1:00 AM on several cloud-free nights in January 2013.

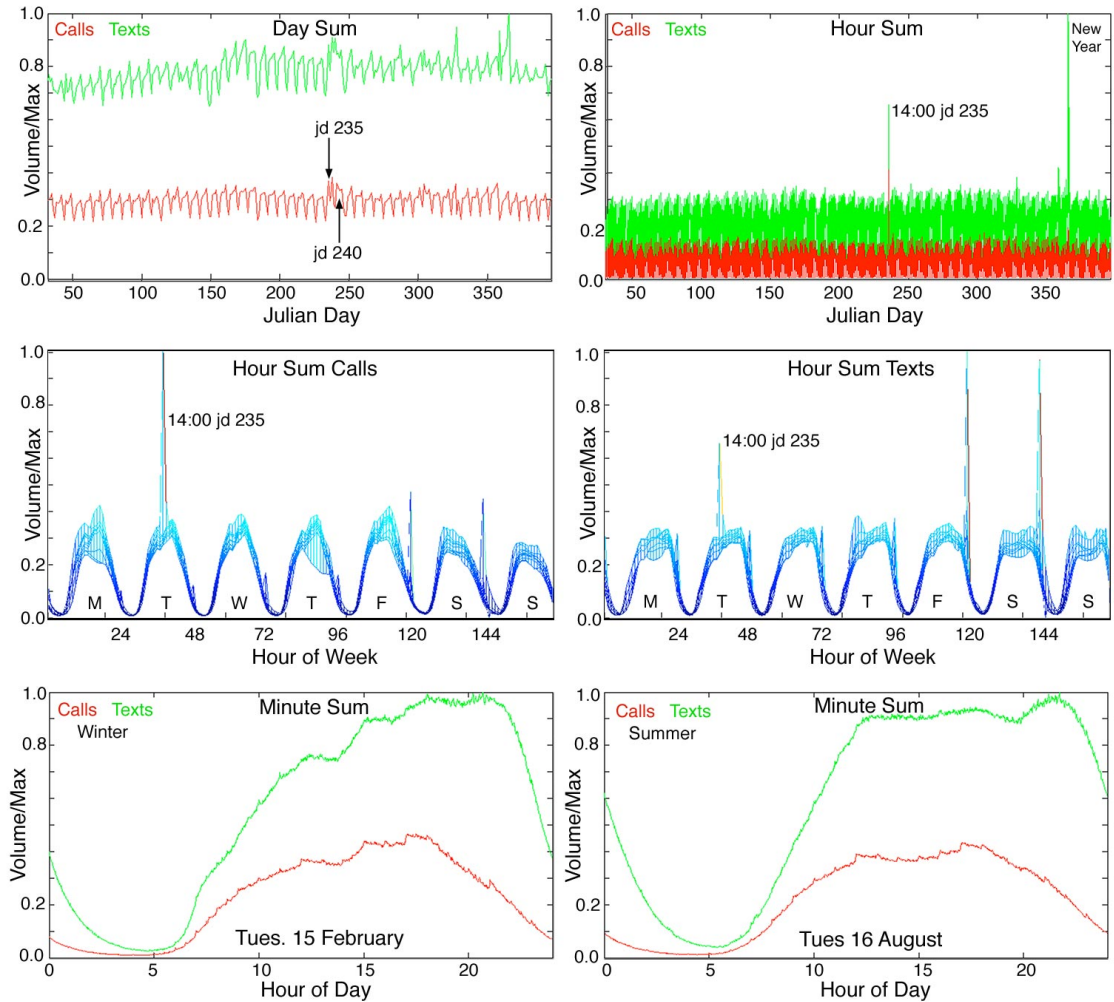

Figure 2 Temporal distribution of mobile communications in the New York metro area in 2011. Spatial averages over ~11,000 sectors show aggregate patterns of $1.5 \times 10^{10}$ voice call starts (calls) and $4.0 \times 10^{10}$ SMS text messages (texts) between 01-02-2011 and 31-01-2012. Weekly and daily cycles differ for calls and texts. Spatial averaging conceals geographic diversity in temporal patterns of individual sectors. Significant disruptions are evident in the hours following the Virginia EQ (13:51 jd 235) and in the days before and after hurricane Irene made landfall in NYC (~0900 jd 240). Spatial average minute sum daily cycles for typical winter and summer weekdays show seasonal differences in texts and calls throughout the day.

Call and text volume density are mapped by bilinear spatial interpolation of activity volumes among centroids of Voronoi sectors. Call and text volumes are aggregated at the scale of angular sectors of Voronoi polygons centered at towers containing multiple antennae whose directional coverage determines the number and azimuths of sectors within each Voronoi polygon. Details of the spatial structure of the sectors are given in the Materials and Methods section and by (Cáceres et al. 2012). The distribution of areas of the ~11,000 sectors in the study area has a median sector area of 4.84 km² with an interquartile range spanning almost two orders of magnitude. To account for geographic differences in antenna density, all temporally aggregated spatial volumes are normalized by sector area to yield call or text density in units of calls or texts per km$^2$ per unit aggregation time (e.g. minute, hour, day). All datasets are gridded using bilinear interpolation with an equal area (UTM projection) grid resolution of 30 m.

The spatial structure of call and text patterns is dominated by strong spatial gradients. The geography of the region results in spatial gradients that vary by orders of magnitude on scales of hundreds of meters in places where sector areas are fine enough to resolve them (e.g. Manhattan). At the spatial scale of the study area, daily call and text volumes are more strongly correlated in space (0.89) than in time (0.70). As a result, even relatively disparate measures of mobile communication, like annual mean call density and annual maximum text density have very similar spatial structure – differing primarily in scale. Comparison of call and text densities with population density and night light brightness shows both similarities and differences related to the quantities being measured and the spatial resolution of each dataset. Comparison of call and text density to population density at the centroid of each sector shows quasi-linear scaling in Log$_{10}$ space over at least 4 orders of magnitude (bottom insets Fig. 1). Scatter about the quasi-linear trend is considerable and asymmetric with heavy upper tail distributions favoring higher call or text volumes at each level of population density. This reflects the fact that at each residential population density there are numerous sectors that experience orders of magnitude higher call and text volumes than would be expected from their residential population density alone. This suggests that many places within the study area have transient (non-resident) populations that contribute to higher call and text volumes than other places with similar resident densities but smaller transient populations. This assertion is self-evident to anyone who has been in midtown Manhattan on a workday. Transient and resident populations in New York have been distinguished in previous studies using by tracking mobility of individual calling locations (Becker et al. 2011), and locality of phone registrations (Girardin et al. 2009).

At the scale of aggregation of the study area, the temporal distribution of spatial mean call and text volumes shows a range of patterns related to the level of temporal aggregation. The plots at the top of Figure 2 compare call and text volumes summed on daily and hourly time intervals. Over the course of the year-long study, daily sums clearly show the weekly cycle and more disruptive events like Hurricane Irene (jd 240) while hourly sums show the nightly drop in call and text activity as well as shorter duration events like the Virginia earthquake (jd 235).

Superimposed plots of hourly call and text volume sums (Fig. 2 middle) for the 52 weeks of the study show both the regularity and variability of daily cycles for each day of the week throughout the year. The prominent spikes in activity (Fig. 2 top) correspond to 2-3pm on the day of the earthquake and to 12-1am on New Years Eve and New Years Day. Spatial averages of 1-minute volume sums (Fig. 2 bottom) of calls and texts for typical winter and summer weekdays illustrate the seasonal similarities and differences in daily cycles.

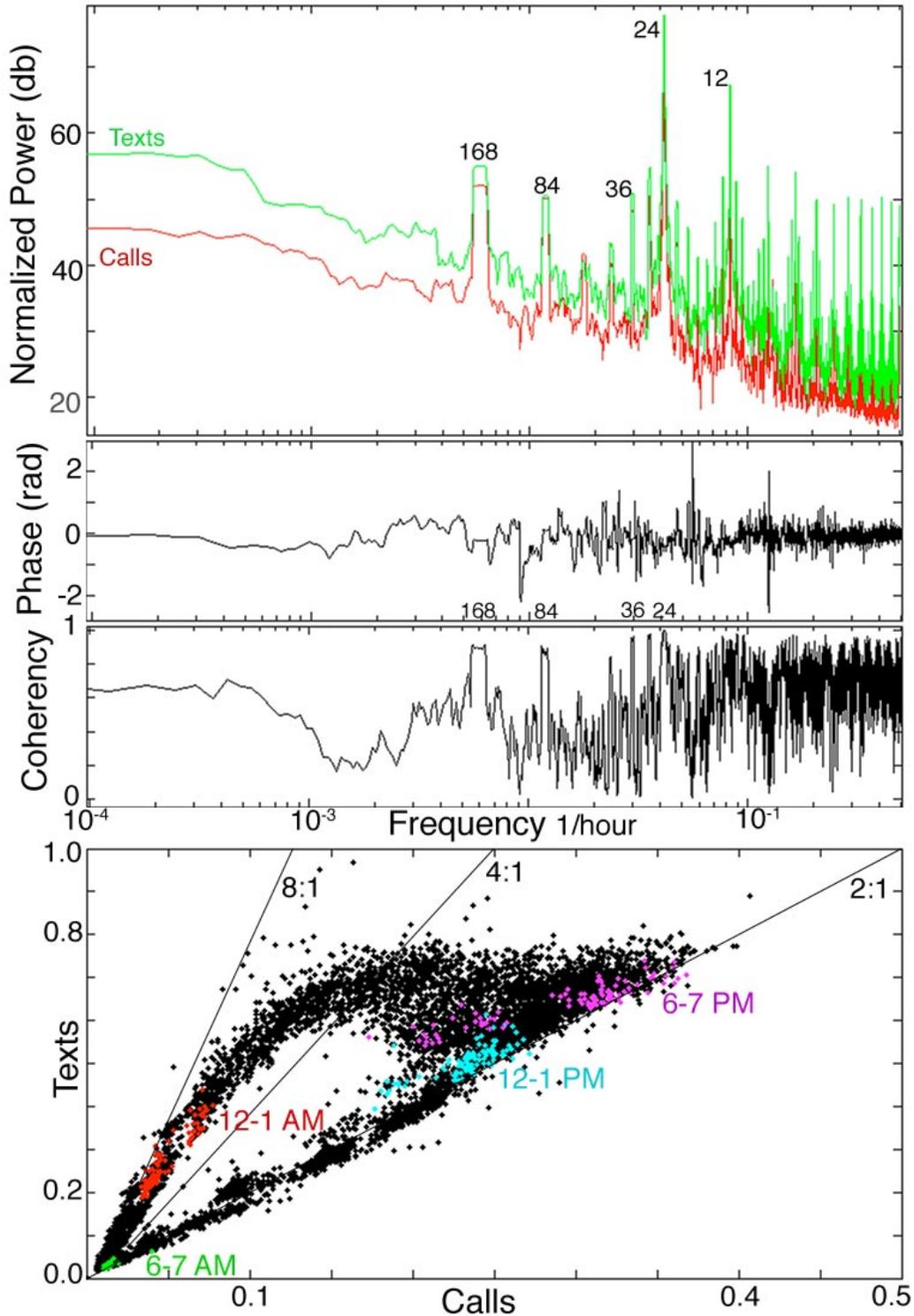

Figure 3 Spectral analysis of cyclicity in calls and texts. Power spectra (top) of spatial mean hourly sums show clear peaks at weekly (168 hours) and daily (24 hour) periods - as well as half-day and half-week periods. Cross spectral analysis (middle) shows high (> 0.8) coherency between calls and texts at these periods, and a slight phase shift in daily cycles. Throughout the year, daytime hourly text and call volumes show a 2:1 ratio persisting until the end of the work day when calls begin to drop while text volumes remain 4x to 8x higher until ~4 AM.

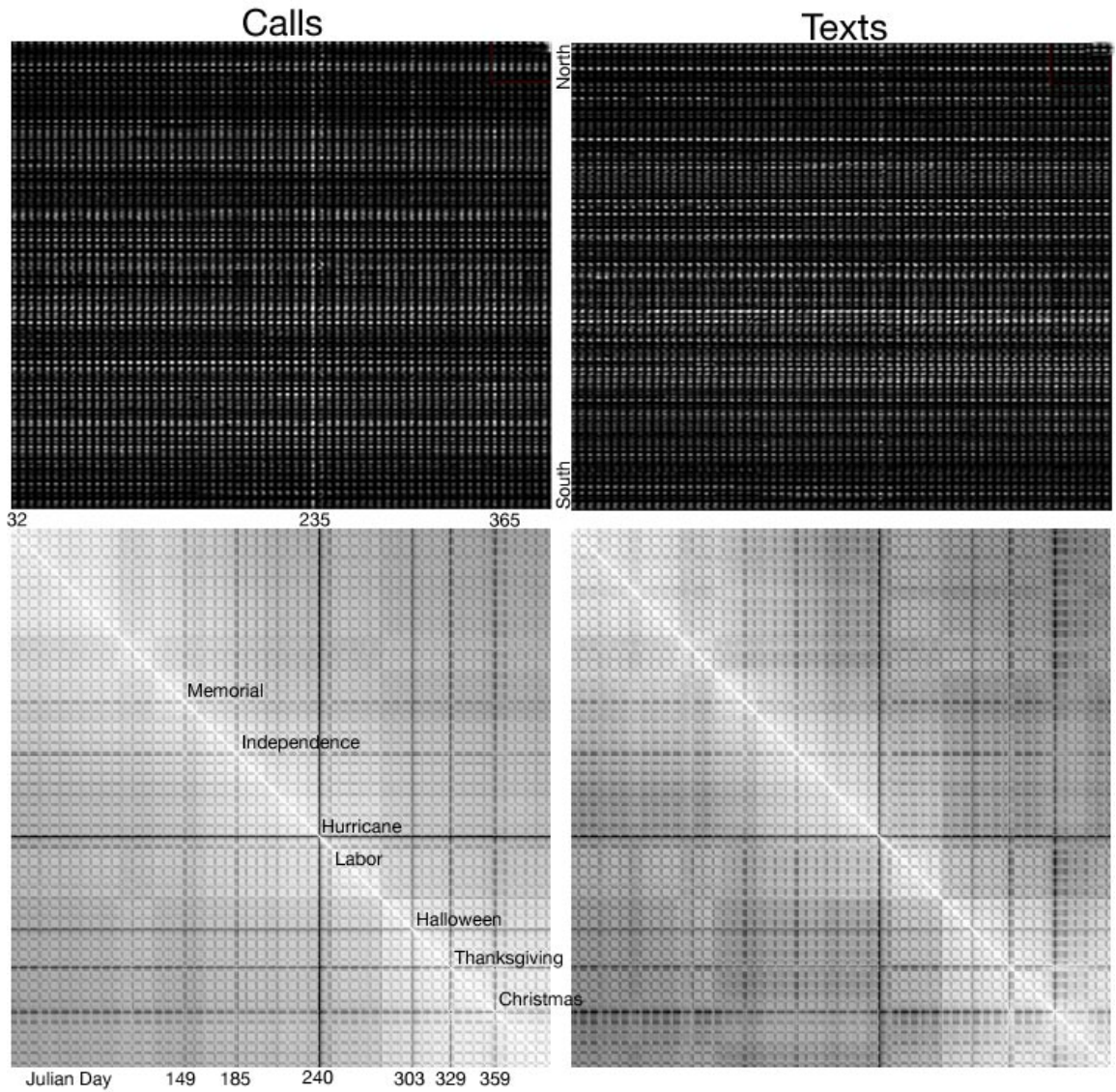

Figure 4 Time-space maps and spatial correlation matrices for calls and texts in the NY metro area in 2011. Each map (top) shows daily volume as a function of time (day) and space (sector latitude). High volume areas appear as lighter horizontal bands. The largest 2% of volumes are saturated white. Increased volumes following the Virginia earthquake appear as thin light bands at jd 235. Each correlation matrix (bottom) shows the spatial correlations of ~11,000 daily volume sums. Lower correlations (darker, black = 0.7) correspond to greater disruptions of the spatial distribution of volumes that characterizes the typical weekly pattern. Note generally lower correlation and abrupt changes in correlation for texts

In addition to the spatial structure resulting from regional geography, the temporal structure of mobile communication patterns is dominated by daily and weekly cycles. We observe call volume cycles similar to those reported in previous studies of New York and other large cities (e.g. (Becker et al. 2011; Reades et al. 2007; Tanahashi et al. 2012) as well as social media (Grauwin et al. 2014). The relative strength of periodicity and relationship between call and text volumes can be quantified with power spectral analysis of the spatially aggregated hourly time series.  Power spectral density estimates (Figure 3 top) of call and text volumes show clear peaks of high variance at daily (24 hour) and weekly (168 hour) periods – as expected.  The power spectra also reveal less obvious half-day and half week periods related to morning and afternoon cycles and Monday-Friday increases in call and text volume. Cross-spectral analysis of call and text volumes shows high coherency (i.e. frequency dependent correlation) at daily and weekly periods as well as half day and half week periods.  The relative phase of call and text periodicities are nearly identical with only slight phase shifts in the daily cycles.  The daily phase shift in call and text volumes is related to the later daily peak in text volumes seen in the bottom panels of Fig. 2.  This phase shift persists throughout the year (Fig. 3 bottom) with text volumes almost exactly twice call volumes throughout the day but remaining high until ~9pm while call volumes peak between 4 and 5pm then drop quickly at the end of the work day.

The temporal periodicity and geographic variability of mobile communication activity can be illustrated efficiently using Time-Space maps in which call and text volumes are shown as a function of Julian day and geographic latitude of sector (Figure 4 top).  At the scale of Fig. 4, the structure of the Time-Space maps is dominated by the weekly cycle and the contrast between higher and lower volume sectors.  The Time-Space map shows all the data without the need for interpolation but some of the spatial and temporal relationships among sectors are not immediately apparent.  Spatial and temporal disruptions to the dominant geographic and cyclic structure are more clearly illustrated using spatial correlation matrices (Fig. 4 bottom).  Spatial correlation matrices show the temporal evolution of spatial patterns.  The spatial correlation among all sectors in the study area changes in time as daily and weekly patterns repeat.  Higher spatial correlations among sector volumes occur at times when spatial distributions of call or text volumes are more similar; lower correlations occur at times when the spatial patterns are less similar.  At the scale of the entire year shown in Fig. 4, the weekly cycle is apparent as the regular grid pattern of higher correlations between pairs of weekdays and lower correlations between weekdays and weekends.  Darker bands of lower spatial correlation occur most prominently on holidays when spatial patterns differ most strongly from the normal patterns occurring throughout the year. The lowest spatial correlations (hence largest spatial disruptions) of the year occur on the days of and preceding the arrival of Hurricane Irene in NYC (jd 240). Note also the more blocky structure and generally lower correlations for text volumes compared to call volumes.  The higher correlations for calls indicate that call volume patterns are more spatiotemporally regular than text volume patterns over the course of the year.  The blockier structure of the text volume correlation matrix suggests that spatial patterns of text volumes change more abruptly throughout the year than spatial patterns of call volumes.

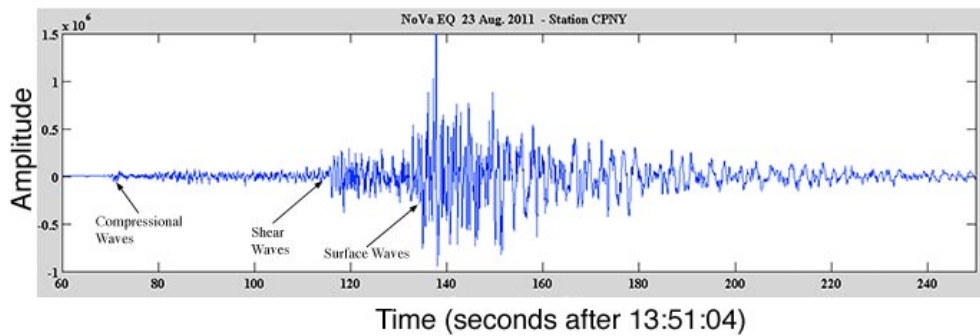
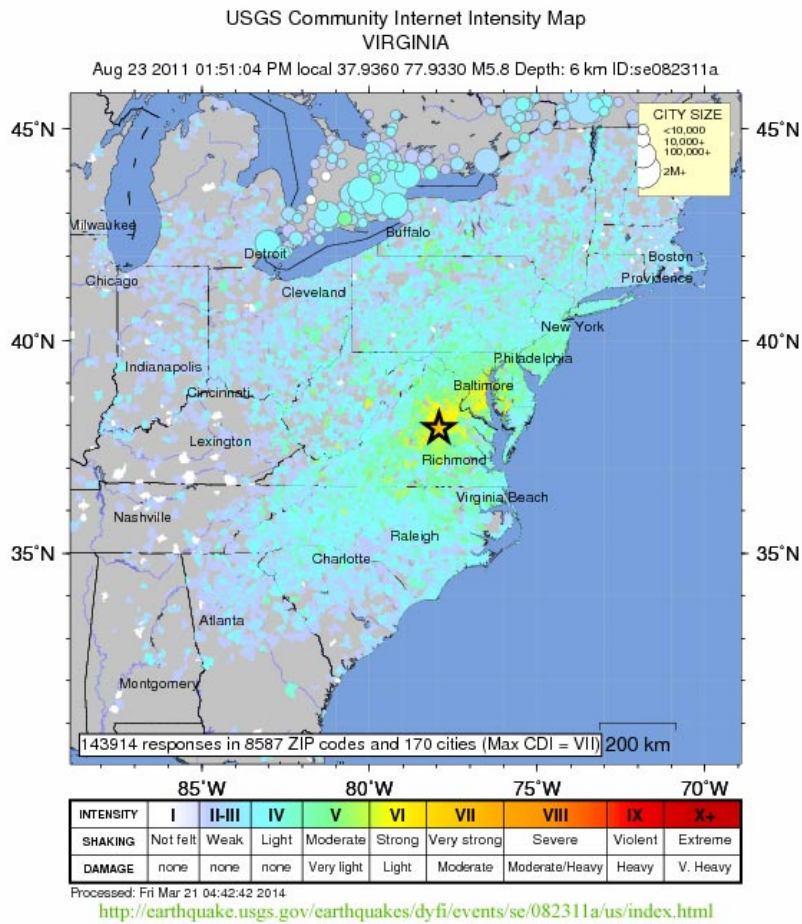

Figure 5 Seismogram and intensity map for the Virginia earthquake of 2011. Seismogram recorded at Central Park in NYC indicates that the first P waves arrived NYC at 13:52 but the much stronger surface waves did not arrive until 13:53. Self-reported intensity map produced the U.S. Geological Survey indicates moderate to light shaking throughout the NY metro area.

*The Virginia Earthquake of 2011*

At 13:51:04 (Eastern Daylight Time) on August 23, 2011 a magnitude 5.8 earthquake occurred at 6 km depth below an epicenter 61 km northeast of Richmond VA. The earthquake was one of the three largest ever recorded in the northeastern US. The US Geological Survey (USGS) received 143,914 intensity reports from observers throughout the eastern US and Canada (Figure 5 bottom). Moderate to light shaking was reported throughout the NY metro area and far beyond. A more detailed account of earthquake-related events and impacts is available at http://en.wikipedia.org/wiki/2011_Virginia_earthquake.

A seismometer located in Central Park in central Manhattan recorded the propagation of seismic waves from the earthquake (Fig. 5 top). While the faster moving compressional waves arrived NYC about a minute after the event, the slower, more powerful surface waves did not arrive until 13:53. The largest seismic waves propagated through the study area in less than one minute but it took approximately three minutes for the full coda of waves to pass through the region. Both call and text volumes abruptly increased throughout the study area at 13:53 – although call volumes increased more rapidly and consistently than text volumes (Figure 6 top/middle). Call and text volume anomalies are defined as the ratio of call (or text) volumes on the day of the earthquake to call (or text) volumes at the same time of the same day one week prior. Call and text volume anomalies are shown as a function of distance from the earthquake epicenter in Fig. 6 (bottom). A spatial distribution of volume anomalies is observed at each distance. Sectors showing larger anomalies than others at the same distance are referred to here as higher response sectors while sectors showing smaller or normal volumes are referred to as lower response. Together, these higher and lower response sectors form the tails of the spatial distribution as it evolves through time. For both call and text volume anomalies, lower response sectors show larger anomalies closer to the epicenter diminishing to smaller anomalies with distance from the epicenter. This is consistent with hemispherical divergence of the seismic wave field's energy density with the expanding wave front. In contrast to the diminishing anomaly of low response sectors with distance from the epicenter, higher response sectors show a slight increase in volume anomaly with distance, reaching a maximum around NYC (470 km) followed by a pronounced decrease at greater distances.

The temporal evolution of the regional response to the earthquake can be quantified by spatial aggregation of call and text volumes at one-minute intervals before, during and after the propagation of the wave field through the study area. Figure 7 (center) shows the mean and standard deviation of the spatial distribution of call and text volumes per minute for the hours preceding and following the earthquake. In the minute the surface waves arrived, between 13:53 and 13:54, the spatial mean ($\mu_C$) and standard deviation ($\sigma_C$) of call volumes increased slightly while the mean ($\mu_T$) and standard deviation ($\sigma_T$) of text volumes dropped. In the following minute both call and text distributions increased abruptly at approximately the same rate.

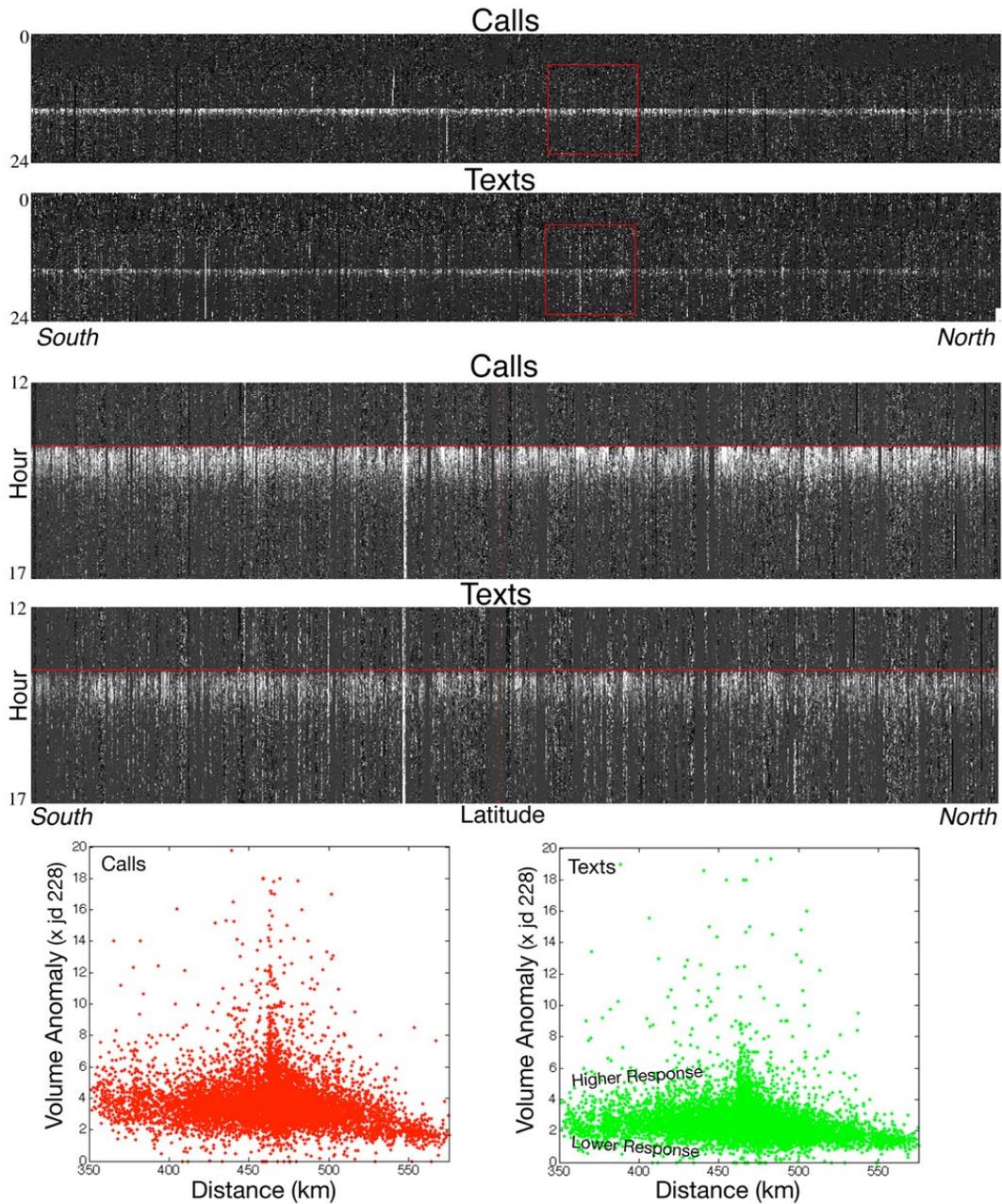

Figure 6 Time-Space maps of calls and texts for the Virginia earthquake. On time-space maps, lighter shade shows higher volumes. Highest 2% of volumes saturated white. Earthquake responses occur throughout the study area (top). Enlargements (red boxes) of sectors near latitude of Central Park (middle) show immediate increases in calls but delayed increases in texts after the arrival of surface waves at 13;53 (red horizontal). Volume anomalies of individual sectors (bottom) show a quasi-linear decrease in anomaly for lower response sectors with distance from the epicenter while higher response sector anomalies increase closer to NYC.

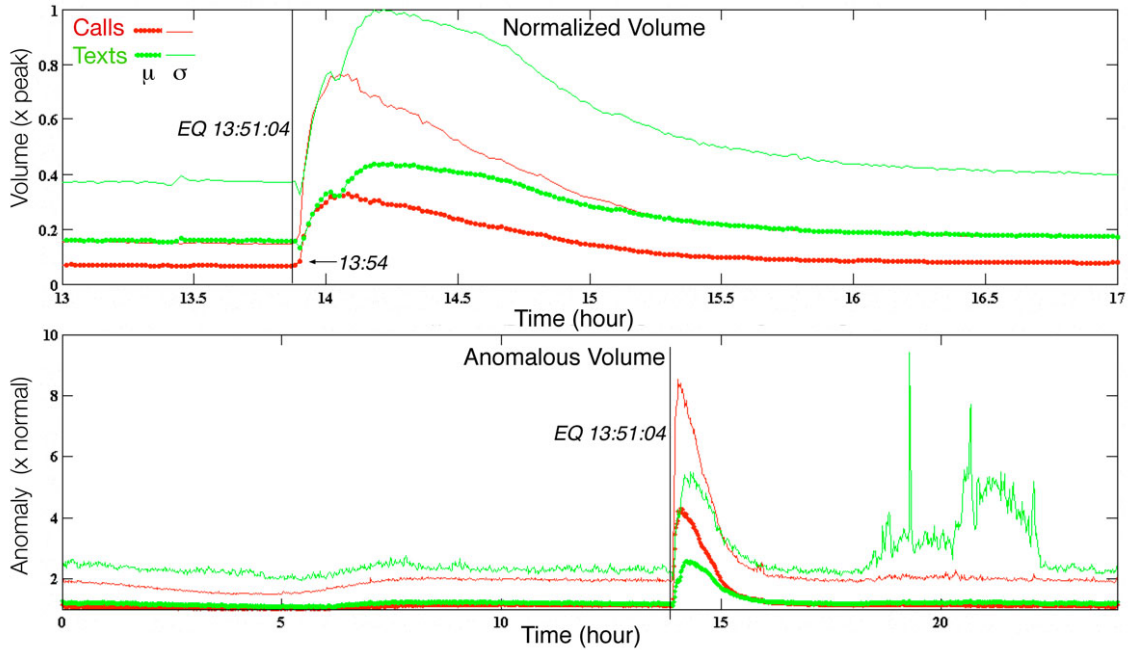

Figure 7 Temporal patterns of mobile communication on 23-08-2011. Volume distributions are shown by spatial means, μ, and standard deviations, σ, of all sectors per minute. Anomalies are volumes normalized by previous week volume to give multiple of normal volume. Maximum call volumes occur ~10 minutes after the earthquake while maximum text volumes are attained ~20 minutes later. Calls decay nearly monotonically after peak but texts decay more slowly before hour 14.7 then decay faster. Note minute to minute spatial variability of evening text volumes.

This monotonic increase in call and text volumes continued for 6 minutes until 14:00, when call volumes reached a maximum plateau and text volumes again dropped slightly. After one minute of little change, text volumes continued to increase for another 10 minutes before reaching a maximum and beginning a monotonic two-stage decrease with a distinct increase in slope around 14.7 hours. After a ~8 minute plateau, call volumes began a monotonic decrease. Following their respective peaks, both call and text volumes took more than one hour to return to pre-earthquake levels.

Distributions of one minute volume anomalies (Fig. 7 bottom), obtained by normalizing mean volumes from the day of the earthquake by volumes from the preceding Tuesday, show a similar pattern – but in terms of anomalous volumes. This plot indicates that mean call volumes increased by more than a factor of 4 relative to normal volumes while mean text volumes increased by less than a factor of 3. Higher response sectors with call volume anomalies one standard deviation greater than the mean showed increases by more than a factor of 8 while text volume anomalies one standard deviation above the mean showed increases by less than a factor of 6. This is consistent with the difference in call and text anomaly distributions seen in Fig. 6 (bottom). Together, these changes indicate that the response of calls to the earthquake was proportionally greater than the response of texts. This is also reflected in the response of maximum volume sectors for calls and texts (Fig. 7 top). This plot shows the highest call or text volume sector in the study area for each minute of the day. The response of the maximum volume sectors for calls is dominated by the earthquake response but the maximum volume sectors for texts show no response to the earthquake and are dominated by three prominent spikes later in the evening and an abrupt increase in the last hour of the day.

The disparity between call and text volume anomalies continues into the evening – hours after the impulse response to the earthquake has decayed back to normal levels. Text volume anomalies show a distinctly different pattern from call volume anomalies later in the evening (Fig. 7 bottom) between ~6pm and ~10pm. While the mean of the text volume anomaly distribution does not change during this time, the standard deviation increases considerably and shows much greater volatility relative to the call volume anomaly distribution. This indicates that the total volume of texts does not change relative to the same times the previous week – but that the sum of the individual sector differences does change considerably from the same times one week prior. This pattern reflects the contrast between the regularity of the spatial distribution of text volumes throughout the work/school day followed by several hours of irregularity (change) in the minute-to-minute spatial distributions of the two days in the evening hours. This spatial variability drops abruptly to normal levels at 10pm, although mean text volumes still do not change. This suggests that the spatial pattern has returned to normal at the end of the day, even while volumes have persisted.

## Hurricane Irene

On August 21, 2011, while passing Puerto Rico, Tropical Storm Irene became the first hurricane of the 2011 Atlantic season. By August 23, Irene had strengthened to a Category 3 hurricane as it passed the Bahamas. On August 27 it weakened to a Category 1 hurricane as it passed the Outer Banks of North Carolina, heading up the US east coast bound for New York. On Friday August 26th the US President and the Governors of New York, New Jersey and Connecticut declared a state of emergency for the New York Metro area and the NYC Mayor's Office issued a mandatory evacuation order for all low lying areas within the five boroughs. Hospitals in these areas were evacuated the same day. Storm surge and high tide predictions for the time of arrival of the hurricane prompted a complete shutdown of the Metropolitan Transit Authority (all subways, commuter trains and busses), private ferries and busses, and all local airports at noon on Saturday, August 27. All Broadway theater shows and the US Open Tennis Championships were cancelled for the following two days. By the time Irene made landfall at Coney Island around 9:00am on Sunday, August 28 it had lost sufficient power to be downgraded to a tropical storm. The hurricane produced a 1.3 m surge coinciding with a 2.9 m tide at Battery Park on the southern tip of Manhattan. Around the time of landfall, the Hudson River flooded evacuation zones A and B in lower Manhattan. Overall, the impact of the hurricane on NYC was less than expected — but it was the first time a hurricane had hit the New York Metro area directly. A more detailed account of hurricane-related effects is available at http://en.wikipedia.org/wiki/Effects_of_Hurricane_Irene_in_New_York.

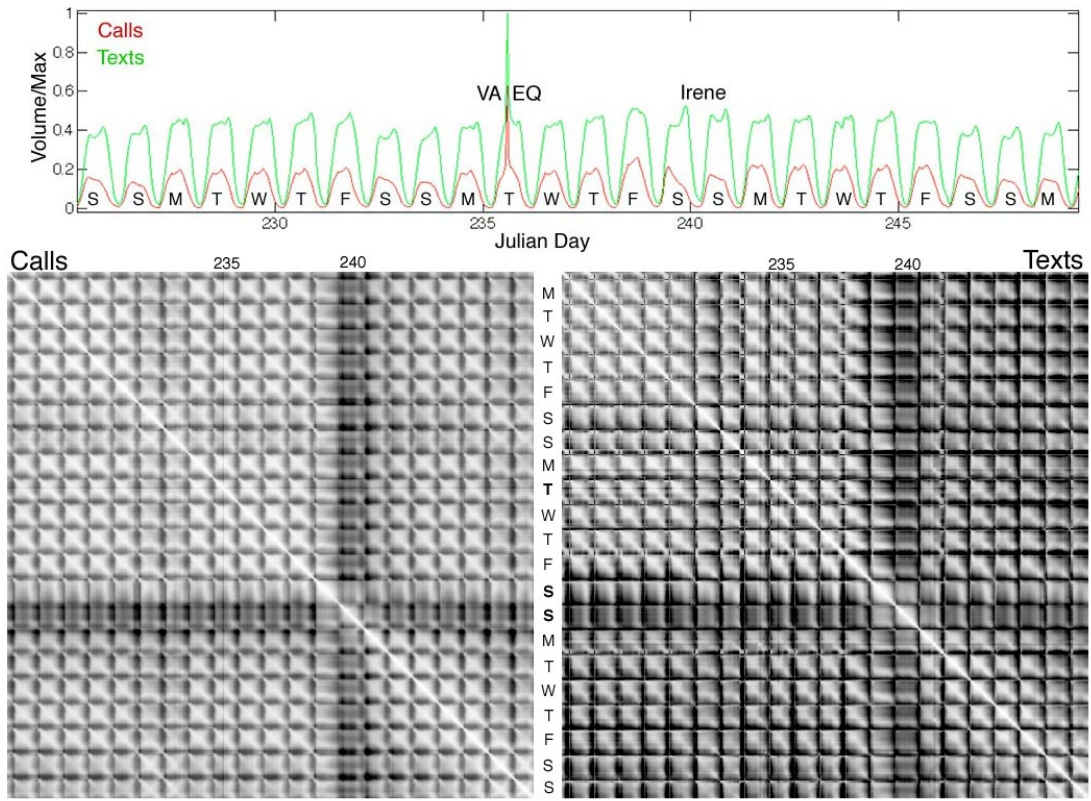

Figure 8 Spatial sum volume time series and spatial correlation matrices for the weeks spanning the earthquake (jd 235) and hurricane (jd 240). Over the region calls decreased and texts increased the afternoon before the hurricane but both quickly returned to normal after the hurricane passed. Hourly correlation matrices show the weekly cyclicity of spatial structure and the disruptions caused by the earthquake and hurricane. Spatial correlation typically peaks at ~0.96 on workdays but dropped briefly to ~0.93 just after the earthquake. Correlations remained high (>0.9) the day before the hurricane indicating a stable spatial pattern that is most strongly correlated with early evening patterns on normal weekdays. The spatial pattern the day of the hurricane shows the weakest overall correlation (< 0.7) to all other days but with its strongest correlation to early evening patterns on weekdays (>0.7).

Unlike the seismic waves from the Virginia earthquake, Hurricane Irene's arrival in NYC was forecast several days in advance. As a result, spatial and temporal patterns in call and text volumes can inform our understanding of both preparation for and response to the hurricane. Specifically, whether and to what extent mandatory evacuations of low-lying coastal areas were heeded or ignored. Spatial and temporal patterns may show both the location and timing of disruptions to the regular communication patterns.

The daily volume time series in Fig. 2 and the correlation matrices in Fig. 4 illustrate the temporal disruption to the regular spatial patterns (or equivalently the spatial disruption to the regular temporal patterns). However, the daily volumes used to compute the correlation matrices in Fig. 4 conceal changes occurring on shorter time scales. Correlation matrices computed from hourly call and text volumes do capture disruptions occurring on time scales of hours before, during and after the hurricane. Figure 8 shows aggregated hourly call and text volumes and their corresponding spatial correlation matrices for the weeks spanning the earthquake and hurricane. At the regional scale of aggregation the spike in volumes following the earthquake is the most conspicuous feature in the time series. Aside from this, the most obvious difference relative to previous or following weeks is the divergence of call and text volumes on the afternoon and evening of the Saturday before the hurricane made landfall. Call volumes were unusually high on Friday PM and Saturday AM but dropped steeply in the early afternoon and precipitously approaching midnight on Saturday night. In contrast, text volumes were unusually high all day on Saturday with a pronounced surge in the afternoon coinciding with the initial drop in call volumes (Fig. 8 top). Text volumes continued to be anomalously high on Sunday during and after the time the hurricane made landfall – despite the fact that call volumes were normal for a Sunday. These temporal patterns complement the information provided by the hourly correlation matrices.

Spatial correlation matrices for call and text volumes provide a compact statistical depiction of how the spatial patterns of volumes evolved through the weeks spanning the hurricane and earthquake. While the earthquake is barely detectable as a brief decorrelation, the hurricane coincides with a prominent band of lower correlation on the jd 239 and jd 240 (Fig. 8 bottom). Superimposed on the 2 day band of lower correlation are hourly variations. During the hour the hurricane made landfall (9-10 am jd 240) the spatial pattern of call volumes shows correlations between 0.45 and 0.72 with other days and times. The temporal variation in spatial correlation of this 9-10 am pattern shows local peaks of ~0.7 twice daily at 8-9 am and 11-12 pm. Correlations for text volumes show a very similar pattern.

Spatial patterns of call and text volume anomalies the day before the hurricane (Figure 9 bottom) show some correspondence to the evacuation zone maps issued by the NYC Office of Emergency Management (Fig. 9 top). Volume anomalies, relative to the Saturday of the previous week, show anomalously high volumes throughout the NYC area with anomalously low volumes on the Atlantic coast of Long Island and in the low lying areas of Manhattan, Brooklyn and Queens and clearly show the airport shutdowns. Areas on the Long Island Sound and upper Hudson River show normal to high volumes. The text volume anomaly map is very similar to the call volume anomaly map. However, the daily volume anomalies conceal changes in the spatiotemporal distributions on shorter time intervals. For a more detailed look at the response to the approaching hurricane, we consider hourly volumes before, during and after the hurricane made landfall.

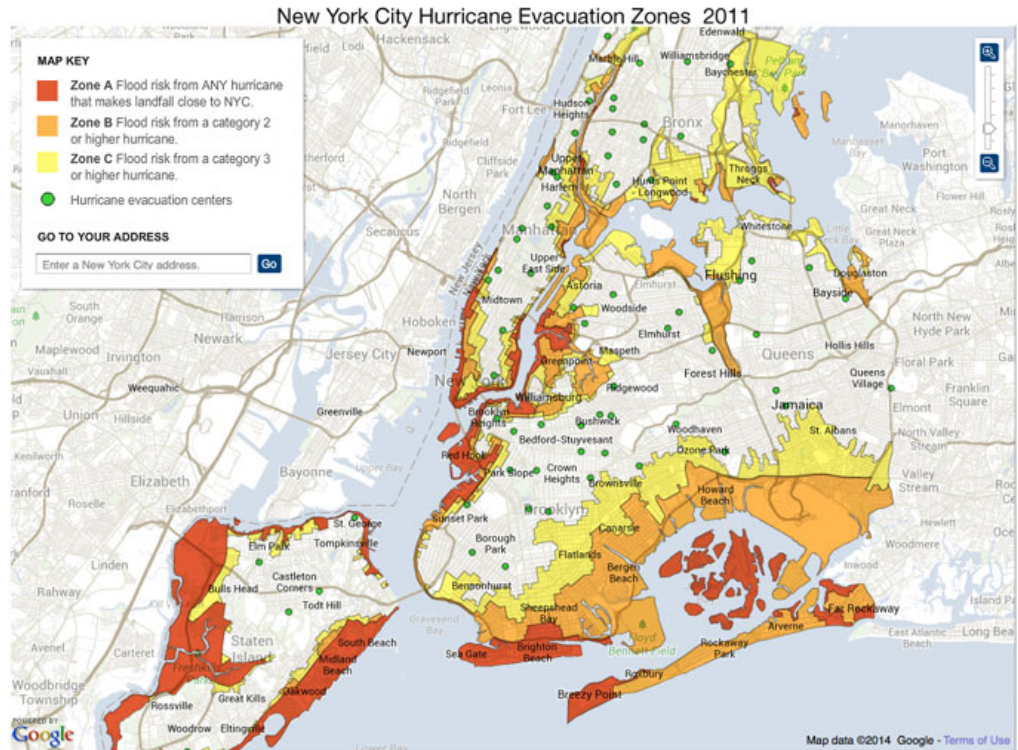
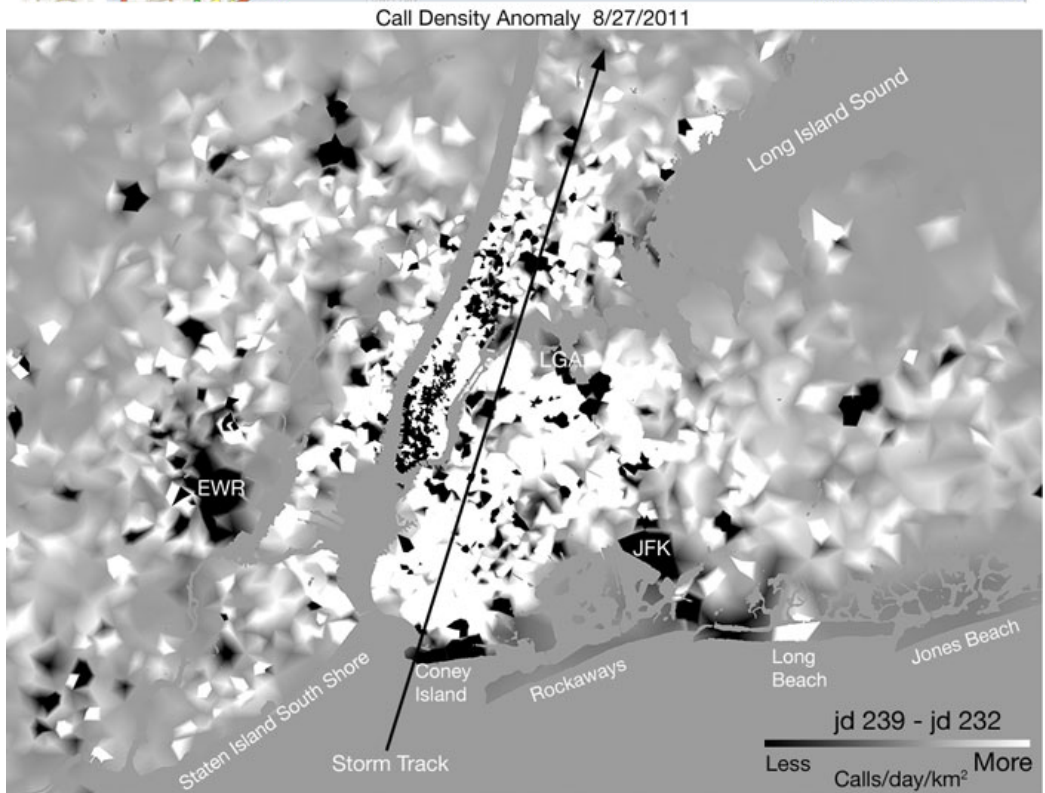

Figure 9 NYC hurricane evacuation zones and call density anomalies. Relative to the previous Saturday, all barrier islands except Long Beach show decreased call volumes the day before the hurricane. Most other evacuation zones also decreased - with some notable exceptions. Overall call density throughout NYC increased relative to the previous Saturday. The spatial pattern for text density shows very similar patterns.

Time series of call volumes from individual sectors in coastal communities show pronounced departures from normal weekend patterns on the weekend of the hurricane. Local airports and public beaches were closed at noon the day before the hurricane made landfall and show abrupt drops in call volume on Saturday afternoon and negligible activity the day of the hurricane. Other coastal communities within the evacuation zones show rapid drops in call volume the afternoon before the hurricane arrived but rapid returns to normal volumes after the hurricane passed on Sunday. Rates and timing of decrease and increase vary somewhat within these areas but most sectors showed call volumes peaking between 1pm and 3pm on the day before the hurricane and declining precipitously after that. This pattern of decrease in call volume is observed at other non-coastal areas but is much more pronounced in the coastal areas shown in Figure 10. In contrast, text volumes show a more variable response throughout the day before the hurricane. Responses range from complete cessation (e.g. Fire Island) to abrupt decrease (e.g. Long Beach) to little change (e.g. Asbury Park) to pronounced increase (e.g. Midland Beach). Given the timing of the hurricane's approach (overnight), text volumes may be more informative than call volumes because the latter normally fall off on Saturday afternoons and evenings.

## Discussion

*Hurricane Response*

Spatial patterns of call and text volume anomalies on the day preceding the arrival of Irene suggest partial, but not full, compliance with evacuation orders for most low lying areas in NYC and surroundings. In most coastal areas call volumes dropped anomalously in the afternoon before the hurricane's arrival, but text volumes showed a much less consistent pattern and often did not decrease in parallel with calls. In terms of total daily volumes, most low lying coastal areas show a decrease in both calls and texts relative to the previous Saturday. However, there are some prominent exceptions, specifically the perplexing increase of call volumes on the barrier island community of Long Beach earlier in the week preceding the hurricane. As yet, we have no explanation for this increase. Despite this preceding increase in volume, call and text volumes in Long Beach did decrease considerably in the afternoon of the day before the hurricane made landfall. Sectors in Evacuation Zone A on the south shore of Staten Island also did not show a decrease in daily call volumes on the day preceding the hurricane. Although daily call volumes did not change appreciably from the preceding week, hourly volumes do show steep drops in the afternoon of the day before the hurricane made landfall. Taken together this suggests that increased volumes earlier in the day offset decreased volumes later in the day to result in a relatively small change in daily volume in these areas.

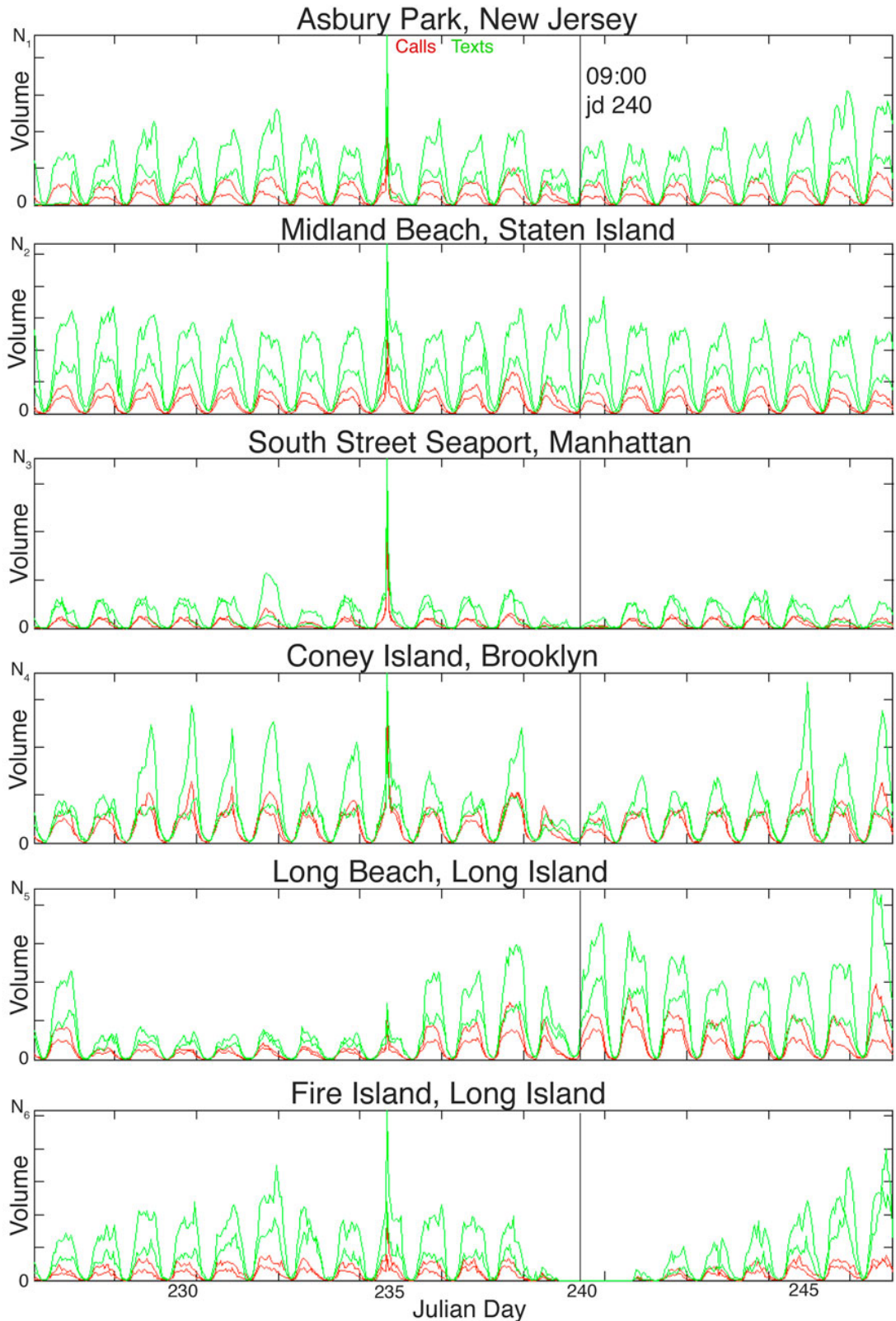

Figure 10 Example time series of call and text volumes for pairs of adjacent sectors in low lying coastal areas for the weeks spanning the earthquake (jd 235) and storm (jd 240). Note difference in call and text volumes the day before the storm.

Also noteworthy is the rate at which call volumes returned to normal in most coastal areas on the morning the hurricane made landfall.  This suggests that evacuees were quick to return to these coastal areas once it was apparent that the impact of the hurricane was generally less than expected.  The mid-morning timing of the hurricane's landfall also introduces some ambiguity in interpretation of the mobile communication patterns.  Because the hurricane's approach and landfall occurred during a phase of the daily and weekly cycle (early Sunday morning) when call volumes are naturally low, the data provide less detailed information about activity immediately before and during the hurricane's landfall.  This highlights a general limitation of mobile communication data.  Interference between the normal daily/weekly cycles of activity and spatiotemporal changes related to other events can mask responses and processes that might be more apparent if the processes occurred during different phases of the cycles.

*Earthquake Response*

The collective response to the earthquake informs our understanding of response to unexpected extreme events in general and earthquakes in particular.  The pervasive and rapid response of high call volumes throughout the study area indicates that the event was widely felt and quickly responded to.   All of the call volume time series that we analyzed clearly show an abrupt increase coinciding with the arrival of the surface waves.  Although some mobile communication outages were reported in areas closer to the epicenter (http://www.washingtonpost.com/business/economy/cellphone-service-falls-short-after-earthquake/2011/08/23/gIQAnl52ZJ_story.html; access 9/11/2014), we find no indication of sector outages (e.g. abrupt drops in volume) or saturation (e.g. constant high volume) in our study area.   The largest aftershock earthquake (magnitude 4.5) occurred at 12:07 AM on 25 August.  We find no disruption in normal mobile communication activity at this time, even in high-density sectors closest to the epicenter (e.g. Trenton NJ).  This suggests that the aftershock was either not widely felt, or did not prompt an unusual response in mobile communication activity.   However, the late hour of the evening at which the aftershock occurred would be expected to reduce response.

The spatial distribution of call volume anomalies is consistent with the expected attenuation of the seismic waves with distance.  The decrease in low response sector volume anomalies with distance from the epicenter (Fig. 6 bottom) suggests that the minimum response to the surface waves did decrease with the amplitude of the waves over the extent of the study area.   The marked decrease in high volume response for sectors more than ~470 km from the epicenter is also consistent with attenuation of the surface waves.  High response (> 5x normal) sectors occur at all but the largest distances (> ~530 km) from the epicenter. The cluster of high response sectors around 470 km corresponds to a large number of small area sectors in midtown and lower Manhattan.  The reason for this clustering requires further investigation.

Because it took the surface waves about a minute to propagate through the study area, the event was effectively instantaneous at the 1-minute resolution of the call and text volume data. The fact that it took over an hour for the call volume anomaly to return to normal suggests that some part of the distribution of callers made multiple calls. The brief drop in text volumes between 14:01 and 14:03 appears to be an aggregate response as it is not observed in individual-sector time series. At the scale of individual sectors, minute-to-minute variability in call and text volumes obscure short duration changes like the brief drop in text volumes. However, the delayed increase in text volumes relative to call volumes is apparent in almost all sector time series that we observed.

The disproportionate call volume anomaly magnitude suggests that the earthquake caused a greater disruption to normal voice communication patterns than to text communication patterns. Immediate call response is apparent in every time series we observed — even in low volume sectors with little or no activity prior to the earthquake. In contrast, for many sector time series of text volumes the response to the earthquake is difficult to detect at all. The disparity between consistent call response and inconsistent text response may be driven by the different demographics of callers and texters combined with geographical variations in these demographic groups. Multiple studies suggest preferential use of SMS over voice communication by younger users in the US (Smith 2011), Norway (Ling et al. 2012) and the Netherlands (Mante and Piris 2002). Despite the different demographics of text and voice communication, and significant variation in demographic distributions from neighborhood to neighborhood in NYC, we consistently observe more rapid increases in calls than texts following the earthquake — even in areas with high (> 4) ratios of texts to calls over the full year.

The combination of immediate call response and delayed text response may result from multi-stage communication patterns in which many people make more calls initially and send more texts later. In many sectors the rate of decrease following the initial surge of calls is similar to the subsequent rate of increase of texts. In comparing number of callers and number of calls per caller, (Bagrow et al. 2011) found that increases in calls following several different extreme events (including an earthquake) were primarily a result of increased number of callers rather than increased numbers of calls per caller. In contrast to the findings of (Bagrow et al. 2011), we find pronounced differences in call and text responses to both events.

In normal, non-event, conditions we also observe several differences between call and text patterns. The most consistent differences are seen in the daily cycles illustrated in Fig. 2. On weekdays, call volumes typically peak in the late afternoon (5-6 pm) while text volumes peak later in the evening (9-10 pm). On weekends, call volumes peak in the early afternoon (1-2 pm) while text volumes have both early afternoon (1-2 pm) and late evening (9-11 pm) peaks. This may reflect both the different uses and different demographics of voice and text communication. This is consistent with the increased spatiotemporal variability of text volumes in evening hours compared to workday patterns.

*Ambient Population Mapping and Modeling*

The spatial structure of call and text density shows strong localization. Much more so than residential population density or night light. Despite strong localization, call and text density maps certainly underrepresent the true degree of spatial clustering of human activities. This is self evident when comparing the meter to decameter spatial scale of the urban fabric (streets, buildings, sidewalks, etc.) with the spatial distribution of antennas and the sectors they define. Although our sector-based geographic distribution of antennas provides considerably greater spatial resolution than tessellations using only tower location without directional information, simple interpolation (bilinear or otherwise) of the relatively coarse hectometer to kilometer scale grid of sector centroids cannot capture the spatial detail of the urban fabric that influences human mobility at scales of meters to decameters. Nonetheless, call and text volume densities do provide a far more detailed depiction of the spatiotemporal distribution of population than survey-based metrics (e.g. census) or remotely sensed proxies (e.g. night light). The strong regularity of daily and weekly cycles in the spatiotemporal distribution of mobile communication suggests that these data may provide a basis for spatially and temporally explicit ambient population models with far greater accuracy and resolution than land cover base proxies derived from remotely sensed imagery. Indeed this has already been investigated by (Isaacman et al. 2012). While mobile communication data do not provide the synoptic coverage of remotely sensed imagery, the two sources are complementary and may ultimately allow for derivation of empirical transfer functions that relate population density distributions to different land use types derived from remotely sensed land cover information.

The analysis presented here uses very simple aggregations, in conjunction with spatial correlation matrices, to quantify the regular spatial and temporal patterns of call and text volumes of the New York Metro area in 2011, as well as the spatiotemporal disruptions that two extreme events caused to these patterns. The diversity of spatial and temporal responses, and the complexity of the structure of the correlation matrices suggest that these data contain far more information than what we have derived from this relatively simple analysis. Preliminary spatiotemporal characterization of the eigenstructure (e.g. (Small 2012)) of the call and text volume covariance matrices distinguish an even greater variety of spatial and temporal patterns – both in the quasi-stationary daily and weekly cyclicities and in the responses to the extreme events. Analysis of these geographically distinctive patterns is the focus of a separate study.

# Materials and Methods

## Spatial mapping of cellular sectors

We use Voronoi tessellation to associate spatial regions with cellular network activity. We group into a sector the set of antennas that reside on the same tower and that share the same compass direction, or azimuth. We use the azimuth information to subdivide the polygons that would result from a tessellation that uses only tower locations. Conceptually, we add edges that bisect the angles between sector azimuths. In practice, we obtain the same result by perturbing the tower locations a small distance in the direction of the sector antennas, then doing the tessellation. Using azimuth information improves the granularity of our spatial mapping of cellular activity. For the New York metropolitan area, the median area of a Voronoi region resulting from our refined tessellation is roughly one quarter of the median area resulting from a tessellation using only tower locations. Additional details on our tessellation method are found in (Cáceres et al. 2012).

## Spatial interpolation of sector centroids

Geographic maps in Figures 1 and 8 are produced by bilinear interpolation among sector centroid locations. Interpolation was done in the Universal Transverse Mercator equal area projection for UTM zone 18N with the WGS84 datum. Volume densities are produced by dividing volumes by sector area (in $km^2$) prior to interpolation. Geographic maps are used for display only; all analyses were done at the scale of individual sector centroids and projected into geographic space as the final step in the process.

## Spectral analysis of volume time series

Power spectra and cross spectra in Figure 3 are derived from spatial averages of hourly volume sums of all sectors over the full study area. Power spectral density, cross-spectral phase and coherency are estimated using the multitaper method (Thomson 1982) with adaptive weighting (Percival and Walden 1993). Multitaper estimation reduces the bias resulting from spectral leakage while minimizing the information loss inherent in the use of conventional tapers and avoiding the need for pre-whitening (Thomson 1982). Adaptive weighting minimizes the mean square error of the spectral estimates by determining the weights for each taper using an iterative procedure that accounts for the (nonwhite) spectral content of the data (Percival and Walden 1993). The cross-spectral coherency provides an estimate of the frequency-dependent correlation between call and text volumes while the cross-spectral phase provides a scale-dependent estimate of any coherent temporal displacement between call and text volumes. Spectral estimates were computed using time-bandwidth products of 2, 4, 8 and 16 and found to be consistent.